\documentclass[fleqn,usenatbib]{mnras}
\usepackage{newtxtext,newtxmath}
\usepackage[T1]{fontenc}

\DeclareRobustCommand{\VAN}[3]{#2}
\let\VANthebibliography\thebibliography
\def\thebibliography{\DeclareRobustCommand{\VAN}[3]{##3}\VANthebibliography}

\usepackage{graphicx}	% Including figure files
\usepackage{amsmath}	% Advanced maths commands

\title[Reconstruction of the dark sectors interaction]{Reconstruction of the dark sectors' interaction: A model-independent inference and forecast from GW standard sirens}

\author[Alexander Bonilla et al.]{
Alexander Bonilla,$^{1}$\thanks{E-mail: abonilla@fisica.ufjf.br}
Suresh Kumar,$^{2}$ $^{3}$\thanks{E-mail: suresh.math@igu.ac.in}
Rafael C. Nunes$^{4}$\thanks{E-mail: rafadcnunes@gmail.com} and
Supriya Pan$^{5}$\thanks{E-mail: supriya.maths@presiuniv.ac.in}
\\
$^{1}$Departamento de F\'isica, Universidade Federal de Juiz de Fora, 36036-330, Juiz de Fora, MG, Brazil\\
$^{2}$Department of Mathematics, Indira Gandhi University, Meerpur, Haryana 122502, India\\
$^{3}$Department of Mathematics, National Institute of Technology, Kurukshetra, Haryana-136119, India\\
$^{4}$Divis\~ao de Astrof\'isica, Instituto Nacional de Pesquisas Espaciais, Avenida dos Astronautas 1758, S\~ao Jos\'e dos Campos, 12227-010, SP, Brazil\\
$^{5}$Department of Mathematics, Presidency University, 86/1 College Street, Kolkata 700073, India
}

\date{Accepted 2022 March 10. Received 2022 March 9; in original form 2021 December 31}

\pubyear{2022}

\begin{document}
\label{firstpage}
\pagerange{\pageref{firstpage}--\pageref{lastpage}}
\maketitle

% Abstract of the paper
\begin{abstract}
Interacting dark matter (DM) - dark energy (DE) models have been intensively investigated in the literature for their ability to fit various data sets as well as to explain some observational tensions persisting within the $\Lambda$CDM cosmology. In this work, we employ the Gaussian processes (GP) algorithm to perform a joint analysis by using the geometrical cosmological probes such as  Cosmic chronometers, Supernova Type Ia, Baryon Acoustic Oscillations, and the H0LiCOW lenses sample to infer a reconstruction of the coupling function between the dark components in a general framework, where the DE can assume a dynamical character via its equation of state. In addition to the joint analysis with these data, we simulate a catalogue with standard siren events from binary neutron star mergers, within the sensitivity predicted by the Einstein Telescope, to reconstruct the dark sector coupling with more accuracy in a robust way. We find that the particular case, where $w = -1$ is fixed on the DE nature, has a statistical preference for an interaction in the dark sector at late times. In the general case, where $w(z)$ is analysed, we find no evidence for such dark coupling, and the predictions are compatible with the $\Lambda$CDM paradigm.  When the mock events of the standard sirens are considered to improve the kernel in GP predictions, we find a preference for an interaction in the dark sector at late times.
\end{abstract}

% Select between one and six entries from the list of approved keywords.
% Don't make up new ones.

\section{Introduction}
\label{sec:intro}

Up-to-date observational data suggest that our universe is mainly driven by a pressure-less or cold dark matter (CDM) and a dark energy (DE) fluid where around 96 per cent ($\sim$\; 28 per cent DM $+$ 68 percent DE) of the total energy budget of the universe is occupied by this joint dark fluid \citep{Aghanim:2018eyx}. The fundamental nature of these fluids, such as, their origin and dynamics are yet to be known even after a series of astronomical missions. Therefore, understanding the dark picture of the universe has remained one of the greatest challenges in cosmology.
In order to reveal the physics of the dark sectors, various cosmological models have been proposed and investigated over the last several years \citep{Copeland:2006wr,Sotiriou:2008rp,Cai:2009zp,DeFelice:2010aj,Capozziello:2011et,Clifton:2011jh,Bamba:2012cp,Cai:2015emx,Nojiri:2017ncd,Bahamonde:2021gfp}. 
The standard cosmological model $\Lambda$CDM is one of the simplest cosmological models which fits excellently to most of the observational probes. However, the physics of the dark fluids is not clear in this model as well $-$ the cosmological constant problem, for instance, is a serious issue \citep{Weinberg:1988cp}. Additionally, in this canonical picture of the universe, 
several anomalies and tensions between different cosmological probes 
may indicate a revision of the $\Lambda$CDM cosmology, see refs. \cite{ 2021arXiv210505208P,Schoneberg:2021qvd}.      
In the $\Lambda$CDM model, we assume the simplest possibility for its ingredients $-$ the independent evolution of DM and DE.  As the physics of the dark sector is not yet clear, there should not be any reason to exclude the possibility of an interaction between these components. By allowing the interaction or the energy exchange between DM and DE, one naturally generalizes the non-interacting scenarios.

The theory of the dark sector interaction did not appear suddenly in the literature. The limitations or issues of the standard cosmological model at the fundamental level motivated to relax the independent evolution of DM and DE. 
For instance, an interaction in the dark sector can provide a possible/promising explanation/solution to the cosmic coincidence problem (\cite{delCampo:2008jx,Velten}), $H_0$ tension \citep{DiValentino:2017iww,Kumar:2017dnp, Yang:2018uae,Pan:2019jqh,DiValentino:2019ffd,Lucca:2020zjb,2021PDU....3300862K,2021CQGra..38o3001D,2021JHEAp..32...28A,2021arXiv211205701R,2021PhRvD.104l3512A,2021Univ....7..300T,DiValentino:2021pow} that arises between the CMB measurements by Planck satellite within the $\Lambda$CDM cosmology \citep{Aghanim:2018eyx} and SH0ES \citep{Riess:2019cxk,2021arXiv211204510R}, and $S_8$ tension \citep{Pourtsidou:2016ico,An:2017crg,Kumar:2019wfs,2021PhRvD.104j4057D,2021PDU....3400899L,2022MNRAS.509.2994A} that arises between the Planck and weak lensing measurements \citep{S8_tension}. Additionally, an interaction in the dark sector could explain the phantom phase of the DE without any need of a scalar field with negative correction \citep{Wang:2005jx,Sadjadi:2006qb,Pan:2014afa,Bonilla_1,Bonilla_2,2021JCAP...10..008Y}. See \cite{Bolotin:2013jpa,Wang:2016lxa} for a comprehensive reading on interacting dark energy models. Therefore, based on such appealing outcomes, it is indeed desirable to consider a wider picture of our universe by including the interaction between DM and DE, and allow the observational data to favor or reject this possibility.

In a standard approach, the interaction between DM and DE is investigated through the inclusion of some phenomenological coupling function to describe the DM and DE dynamics intuitively. However, let us recall that some action formalisms, i.e., construction of the DE-DM interaction models from the first principle, including the Noether symmetry approach, have also been developed in the literature, see e.g. \cite{2020PDU....2700444P,Gleyzes:2015pma,Boehmer:2015kta,Amico:2016qft,Kase:2019hor,2018arXiv180900556V,Pan:2020zza}. On the other hand, given the great interest of the community in this theoretical framework, accomplishing a model independent analysis becomes a necessary task. In principle, one may do it using cosmographic approach, wherein a series expansion is performed around $z = 0$ for a cosmological observable, and then the data are used to constrain the kinematic parameters. This procedure works fine for lower values of $z$, but may not be good enough for larger values of $z$, see \cite{2021PhRvD.104l3518L}. An interesting and robust alternative could be to consider a Gaussian process (GP) to reconstruct the cosmological parameters in a model-independent way \citep{GP_01,GP_02,GP_03,GP_04,GP_05,GP_06,GP_07,GP_08,Jesus2020,GP_09,GP_10,2021arXiv211014950M,2021ApJ...915..123S,Bernardo:2021cxi,Dialektopoulos:2021wde,Bengaly:2021wgc,Avila:2022xad}
or to fix a class of cosmological models \citep{2020CQGra..38e5007B,2021JCAP...09..014B,2021JCAP...07..048R,2021arXiv210401077E,2021PDU....3200812R}. The GP and other alternative approaches have been applied to reconstruct an interaction between DM and DE in a minimally model-dependent way in various works with different data sets and approximations \citep{Yang2015, Wang2015, GP_IDE_01, GP_IDE_02, GP_IDE_03, GP_IDE_04, GP_IDE_05, GP_IDE_06, GP_IDE_07}.

In this work, we employ the GP to carry out a joint analysis by using some geometrical cosmological probes, viz.,  Cosmic chronometers (CC), Supernova Type Ia (SN), Baryon Acoustic Oscillations (BAO), and the H0LiCOW lenses sample to constrain/reconstruct the interaction in the dark sector of the universe in two different frameworks, namely, the one where the EoS of DE mimics the vacuum energy (known as an interacting vacuum energy scenario) and secondly a general coupling scenario where DE is allowed to assume a dynamical character via its equation of state (EoS). This latter possibility has not been studied much in the literature, viz., most of the works are carried out with only the constant or linear approximation of the EoS parameter of DE. Moreover, to our knowledge of the current literature, the reconstruction of the interaction in the dark sector has not been performed using a joint analysis. In addition, we also simulate a catalogue of 1000 standard siren events from binary neutron star mergers, within the sensitivity predicted for the third generation of the ground GW detector called the Einstein Telescope (ET), and we use these mock data to improve the reconstruction of the coupling function from the SN, BAO, CC and H0LiCOW data. A model-independent joint analysis from above-mentioned data sets, including a forecast analysis with the simulated data for optimizing the covariance function (or kernel in GP language), as we present here, to our knowledge is new and not previously investigated in the literature. Indeed, a joint analysis with several observational probes is helpful to obtain tight constraints on the cosmological parameters. In this work, we develop this methodology to obtain an accurate and robust reconstruction of a possible interaction between DM and DE.

The paper is structured as follows. In Section \ref{sec-method-data-theory}, we describe the GP, the observational data sets and the theoretical framework used in this work for model-independent inference of the dark sector coupling. In Section \ref{sec-results},  we present and discuss our results on the reconstruction of the coupling function between DM and DE following the model-independent approach, wherein the subsections \ref{sec-ivs} and \ref{sec-ide} describe two different reconstructed scenarios. Further, in Section \ref{sec-gw}, we use the mock gravitational waves data in order to get a more deeper understanding on the evolution of the coupling function. Finally, in Section \ref{sec-conclu}, we conclude our work with a brief summary of the entire study.

\section{Methodology, data sets and the theoretical background}
\label{sec-method-data-theory}

This section is divided into the following three parts: the Gaussian process, the observational data, and a basic framework of the theory, which we are going to test in this article using the observational data following the model-independent Gaussian approach.

\subsection{Gaussian process}
\label{sec-gaussian}

In a nutshell, the GP in cosmology allows us, given an observational data set $f(z)\pm \sigma_f$, to obtain a function $f(z)$ without the need to assume a parametrization or physical model about the dark nature of the main components of the universe. The GP method adequately describes the observed data based on a distribution over functions. The reconstructed function $f(z)$ (and its derivatives $f'(z)'$, $f''(z)$,..., etc) have a Gaussian distribution with mean and Gaussian error at each data point $z$. The functions at different points $z$ and $z'$ are related by a covariance function $k(z,z')$, which only depends on a set of kernels with hyperparameters $l$ and $\sigma_f$, describing the strength and extent of the correlations among the reconstructed data points, respectively. Thus, $l$ gives a measure of the coherence length of the correlation in the x-direction, and $\sigma_f$ denotes the overall amplitude of the correlation in the y-direction. In general, the hyperparameters are constant since their values point to a good fit of the function rather than a model that mimics this behavior, which means that the GP optimizes both concerning the observed data. Finally, the GP method is model-independent in a physical model and assumes a particular statistical kernel that determines the correlation between the reconstructed data points. The entire methodology used in this work is described in detail in section II of Ref.~\cite{Bonilla:2020wbn}.

\subsection{Observational data sets}
\label{sec-data}

In this section we shall describe the geometrical probes in detail that we have used to trace the interaction in the dark sector.

\begin{itemize}

\item Cosmic Chronometers (CC): The CC approach is  very powerful to detect the expansion history of the universe that comes  through the measurements of the Hubble parameter. Here we take into consideration 30 measurements of the Hubble parameter distributed over a redshift interval $0 <  z <  2$ as in Ref. \cite{Moresco16}. \\

\item Supernovae Type Ia (SNs): The first astronomical data probing the accelerating expansion of our universe are SNs. Certainly, SNs  are very important astronomical probes in analysing the properties of DE and the expansion history of the universe. The latest compilation of SN data  (Pantheon sample) that we have used in this work, consists of 1048 SN data points in the redshift  range $0.01 <  z <  2.3$ \citep{Scolnic18}. In the context of a universe with zero curvature, the entire Pantheon sample can be summarized in terms of six model-independent $E(z)^{-1}$ data points \citep{Riess18}. Here we use the six data points as reported in ref. \cite{Haridasu18} in the form of $E(z)$ taking into account the theoretical and statistical considerations for its implementation. \\

\item Baryon Acoustic Oscillations (BAO): Another important cosmological probe is the BAO data. With the use of BAO, the expanding spherical wave produced by baryonic perturbations of acoustic oscillations in the recombination epoch can be traced through the correlation function of the large-scale structure displaying a peak around 150$h^{-1} {\rm Mpc}$. Here we have used BAO measurements from various astronomical surveys: (i) measurements from the Sloan Digital Sky Survey (SDSS) III DR-12 which report three effective binned redshifts $z = 0.38, 0.51$ and $0.61$ \citep{Alam17}, (ii) measurements from the clustering of the SDSS-IV extended Baryon Oscillation Spectroscopic Survey DR14 quasar sample reporting four effective binned redshifts $z = 0.98, 1.23, 1,52$ and $1.94$, as in \cite{Zhao19}, (iii) measurements from the high-redshift Lyman-$\alpha$ survey reporting two effective binned redshifts  at $z = 2.33$ \cite{du_Mas20}  and $z = 2.4$  \cite{du_Mas17}. All the measurements are presented in terms of $H(z) \times (r_d/r_{d,fid})$ km s$^{-1}$Mpc$^{-1}$, where $r_d$ denotes the co-moving sound horizon and $r_{d,fid}$ is the fiducial input value provided in the above surveys. \\

\item H0LiCOW sample: Finally, we use the sample from the $H_0$ Lenses in COSMOGRAIL's Wellspring program \footnote{\url{www.h0licow.org}}, another geometrical probe in this list which measures the Hubble constant in a direct way (without assuming any model in the background). The H0LiCOW collaboration has measured six lens systems, determined by measurements of time-delay distances, $D_{\Delta t}$, between multiple images of strong gravitational lens systems due to elliptical galaxies \citep{H0LiCOW}. The entire information is encapsulated in the time-delay distance $D_{\Delta t}$.  Along with these six systems of strongly lensed quasars, the angular diameter distance to the lens $D_l$ also offers some additional information in terms of four more data points. Therefore, in total, one can employ 10 data points and we have used them in this work (we refer to \cite{Birrer2019,Pandey2020} for more details in this context).  

\end{itemize}

\subsection{Theoretical framework}
\label{sec-theory}

For a model-independent theoretical description of the dark sectors' interaction, in this work, we follow the similar methodology as in ref. \cite{Yang2015}. In the context of a Friedmann$-$Lema\^{i}tre$-$Robertson$-$Walker universe, we assume that the total energy density of the universe is comprised by DE and DM only where both of them are coupled through a non-gravitational interaction.  Thus, the conservation equations for DM and DE are modified as 

\begin{eqnarray}
&&\dot{\rho}_{\rm DM} +3 H \rho_{\rm DM}  = -Q (t)~,\label{cont1}\\
&&\dot{\rho}_{\rm DE} + 3 H \rho_{\rm DE} (1+w)= Q (t)~,\label{cont2}
\end{eqnarray}

where $w  = p_{\rm DE}/\rho_{\rm DE}$ is the equation of state of DE ($p_{\rm DE}$ denotes the pressure of the  DE fluid), $H=\dot{a}/a$ is the expansion rate of the universe and it is related to the total energy density of the universe as $3H^2 = \rho_{\rm DM} + \rho_{\rm DE}$ (in the units where $8 \pi G = 1$). The function $Q (t)$ describes the interaction between DM and DE, and usually it is taken to be a function of the energy densities of DM and DE. For $Q (t) = 0$ with $w =-1$, the standard $\Lambda$CDM cosmology is recovered. Now, combining the conservation  equations (\ref{cont1}) and (\ref{cont2}) with expansion  rate of the universe $H(z)$, we obtain \citep{Yang2015}:

\begin{eqnarray}
\label{eqn:WqE}
-wq  &=& 2 \Big(E E'^2 + E^2 E'' - \frac{w'}{w} E^2 E' \Big) (1+z)^2\nonumber \\ 
&&- \Big[ 2(5 + 3 w)E^2 E' - 3 \frac{w'}{w} E^3\Big](1+z)\nonumber \\
&&+ 9(1 + w)E^3,
\end{eqnarray}  

where for convenience, we have used a dimension-less variable $q = Q (t)/H^3_0$ to characterize the interaction,  $E(z)=H(z)/H_0$ is the normalized Hubble rate and the prime denotes the differentiation with respect to the redshift $z$. Let us note that the symbol $q$ is usually used to represent the deceleration parameter in the literature, but here this symbol has a different meaning as defined above.  The detailed derivation of equation (\ref{eqn:WqE}) is given in Appendix \ref{sec-appendix}. Now, using the normalized co-moving distance,

\begin{eqnarray}
\label{eqn:D}
D = \frac{H_0}{c} \left(\frac{1}{1+z} \right) d_L(z),
\end{eqnarray}  
where $d_L(z)$ represents the luminosity distance at redshift $z$, eq.(\ref{eqn:WqE}) can be expressed alternatively as 

\begin{eqnarray}
\label{eqn:WqD}
-wq  &=& 2 \Big(\frac{3 D''^2}{D'^5}  - \frac{D'''}{D'^4} + \frac{w' D''}{w D'^4} \Big) (1+z)^2\nonumber \\ 
&& + \Big[2(5 + 3w)\frac{D''}{D'^4} + \frac{3 w'}{w D'^3}\Big](1+z)\nonumber\\
&& + \frac{9(1 + w)}{D'^3}.
\end{eqnarray}

The above methodology represents a general framework to reconstruct the coupling function with minimal assumption. The only assumption of fact is the validity of the cosmological principle, and a possible coupling between DM and DE has been assumed as a theoretical prior, where this second assumption must be tested with the observational data. 

In what follows, we will test this theoretical framework.

\begin{figure*}
\begin{center}
\includegraphics[width=3.1in]{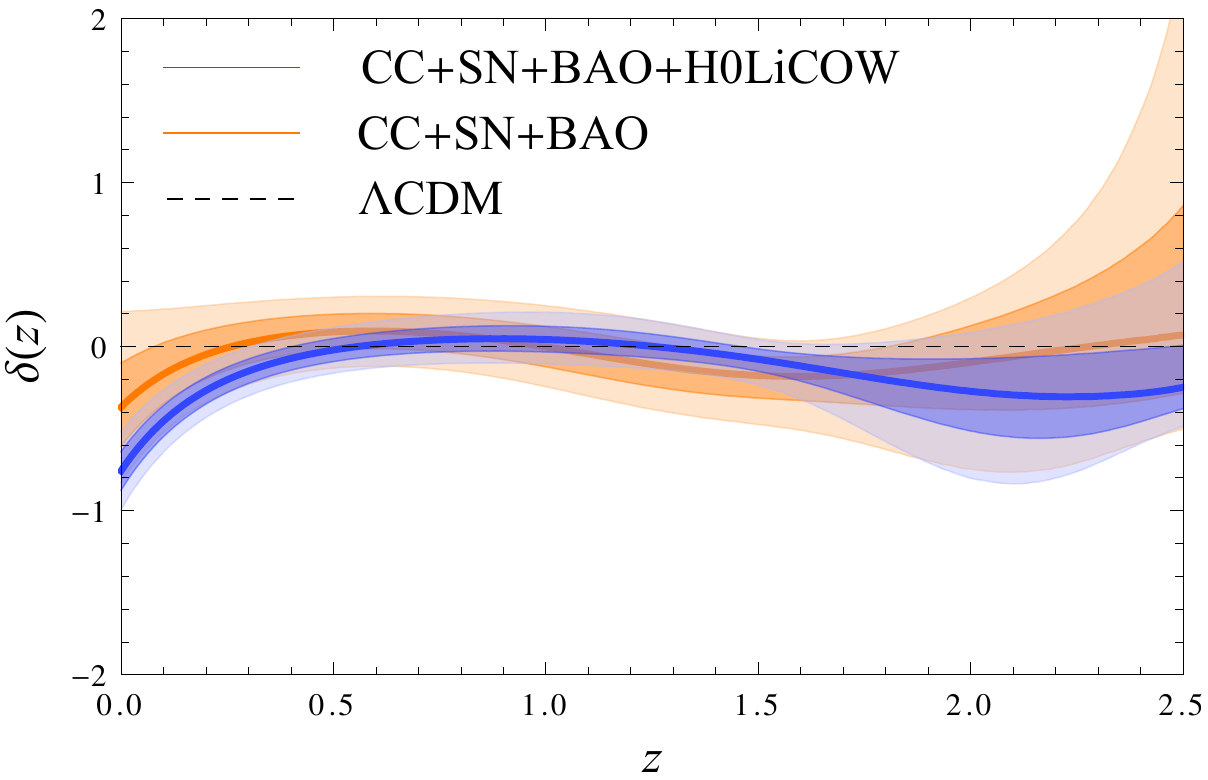} \,\,\,\,
\includegraphics[width=3.1in]{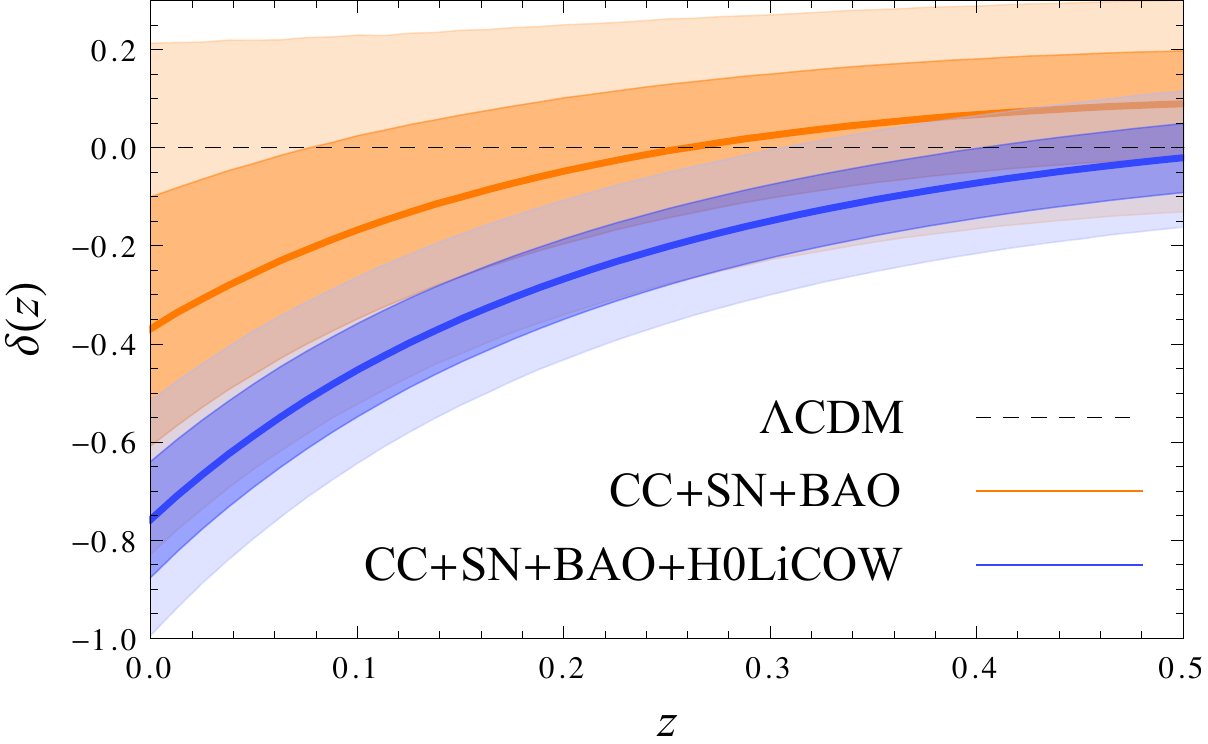}
\caption{Left-hand panel: Reconstructed coupling function $\delta (z)$ at $1\sigma$ and $2\sigma$ CL in the interacting vacuum energy scenario from CC+SN+BAO (Orange) and CC+SN+BAO+H0LiCOW (Blue) data. Right-hand panel: The same as in left-hand panel, but restricted to the range $z \in [0, 0.5]$. The dashed black curve corresponds to the canonical $\Lambda$CDM prediction and the solid curves represent the GP mean. }
\label{IVCDM_results01}
\end{center}
\end{figure*}

\begin{figure*}
\begin{center}
\includegraphics[width=3.1in]{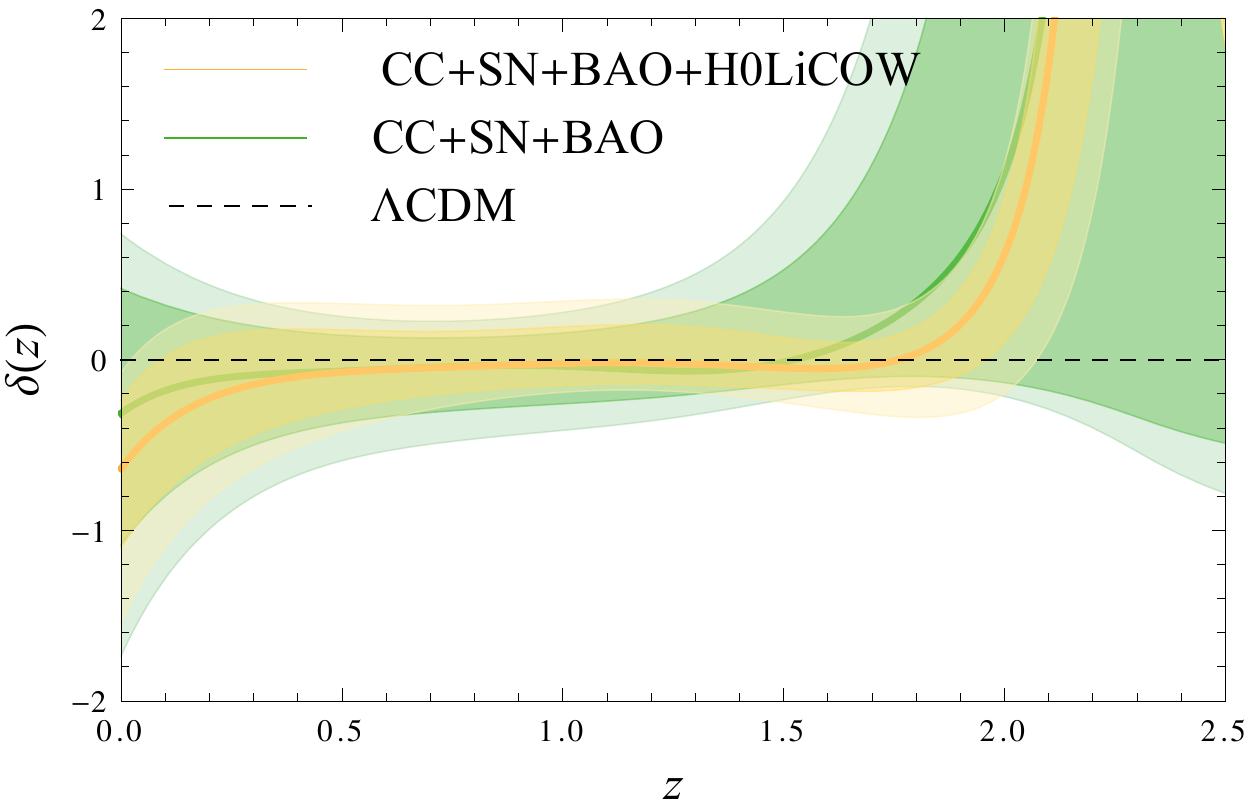} \,\,\,\,
\includegraphics[width=3.1in]{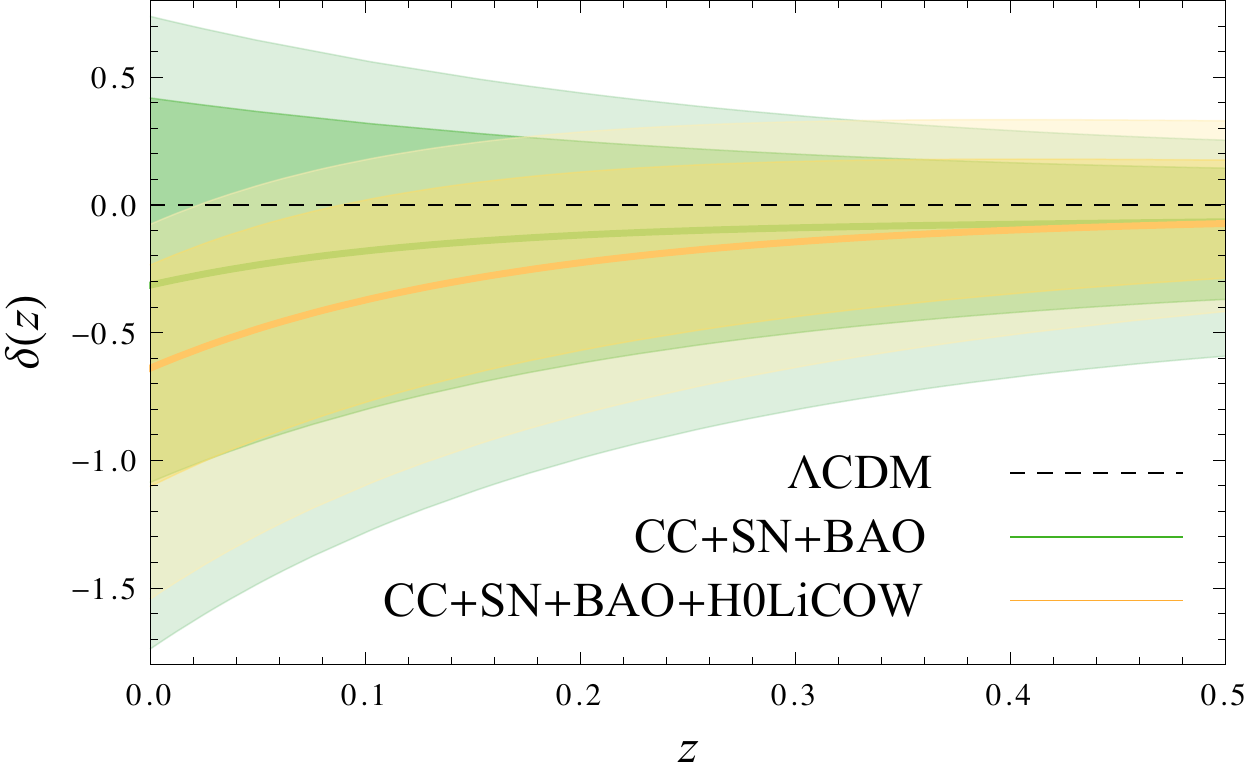}
\caption{Left-hand panel: Reconstructed coupling function $\delta (z)$ at $1\sigma$ and $2\sigma$ CL in the general interacting scenario of the dark sectors' from CC+SN+BAO (Green) and CC+SN+BAO+H0LiCOW (Yellow) data. Right-hand panel: The same as in left-hand panel, but restricted to the range $z \in [0, 0.5]$. The dashed black curve corresponds to canonical $\Lambda$CDM prediction and the solid curves stand for the GP mean. }
\label{Geral_results}
\end{center}
\end{figure*}

\section{Results and Discussions}
\label{sec-results}

In this section, we present and discuss the results of our analyses considering two separate cases.  First, we fix the EoS of DE to $w=-1$, and reconstruct the interaction function $q = Q (t)/H_0^3$. This possibility characterizes a very well-known sub-class of the interaction scenario in the dark sector, known by interacting vacuum energy. Secondly,  we consider a very general possibility assuming $w$ as a free and dynamical function, and similarly reconstruct the interaction function $q$. Thus, having both the possibilities, we explore a very general description of the dark coupling in a model-independent approach. Before we enter into the main results, we rescale the function $q$ (see eq.(\ref{eqn:WqD})) to $\delta (z) = q(1+z)^{-6}$.  Such a pre-factor is just considered as a scale transformation with respect to $z$, introduced to better expose the results in the graphical description. Thus, here onwards the function $\delta (z)$ will characterize the coupling function.

We proceed further considering the above two scenarios of coupling in the dark sector.  To reconstruct $\delta (z)$, we use $M_{9/2}$ kernel in all the analyses performed in this work. In the case where we assume $w(z)$ to be a free function, we follow the same methodology as presented in \cite{Bonilla:2020wbn}. For this purpose, we have used modified versions of  some numerical routines available in the public  GAPP (Gaussian Processes in Python) code \cite{GP_01}. In all of our analyses, we employ GP to perform a joint analysis using the minimal data set combination CC+SN+BAO, which to our knowledge has not been investigated previously  in the literature. We now present and discuss our main results.

\subsection{Interacting Vacuum Energy}
\label{sec-ivs}

In Fig. \ref{IVCDM_results01}, we have shown the reconstruction of $\delta(z)$ using the data combinations CC+SN+BAO and CC+SN+BAO+H0LiCOW. In both analyses, we note that for $z > 0.5$ the dynamical coupling function $\delta (z)$ between the dark components is statistically well compatible with $\delta (z) =0$. It is interesting to note that the GP mean predicts a possible oscillation in $\delta (z)$, where we can note an oscillation between positive and negative values in the analysed range of $z$. This result strengthens some earlier interaction models having a sign changeable property, see for instance \cite{Pan:2019jqh,Pan:2020bur,2021PhRvD.103h3520Y}.  
For the present scenario, at late cosmic time, i.e. for $z < 0.5$ ($z < 0.25$),  we find a trend towards $\delta < 0$ for  CC+SN+BAO+H0LiCOW (CC+SN+BAO) data.  When evaluated at present moment, we find $\delta(z=0) = -0.37 \pm 0.24$ ($-0.76 \pm 0.12$) at $1\sigma$ CL from CC+SN+BAO (CC+SN+BAO+H0LiCOW) data. This suggests an interaction in the dark sector at more than $3\sigma$ CL from the CC+SN+BAO+H0LiCOW joint analysis. It is important to emphasize that these constraints on $\delta(z)$ are subject to the condition $w=-1$. Also, we notice that the combined analysis with several data sets offers a more stringent bound on the interaction function compared to \cite{Yang2015},  where only the SN Ia Union 2.1 data set~\cite{Suzuki:2011hu} was employed.

\subsection{General interaction scenario in the dark sector}
\label{sec-ide}

In the previous subsection, we analysed a particular interaction case, namely the interacting vacuum-energy ($w=-1$) to get the constrains on $\delta(z)$. As a second round of analysis, we relax these conditions by assuming $w(z)$ to be a free function. This possibility allows us to reconstruct the coupling in the dark sector in a general way, because in this case, no physical assumption is considered on the EoS of DE.    
In Fig. \ref{Geral_results}, we show the reconstruction of $\delta(z)$  from CC+SN+BAO and CC+SN+BAO+H0LiCOW data combinations. Since in this scenario, we have an additional free parameter $w$ to propagate errors compared to the previous case with $w =-1$, it is expected  that large error bars might be imposed on the reconstructed $\delta (z)$  compared to the case with $w=-1$. 
As a general feature of the GP mean, we can note a flux of energy from DM to the DE  at high $z$, and as cosmic time evolves up to approximately $z < 0.5$, the coupling function $\delta (z)$ reverses its sign, at low $z$ (late time). This again goes in support of some phenomenological models of the interaction \citep{Pan:2019jqh,Pan:2020bur}. In this general framework, we find $\delta(z=0) = -0.31 \pm 0.77$ at $1\sigma$ CL from CC+SN+BAO data and $\delta(z=0) = -0.64 \pm 0.43$ at $1\sigma$ CL from CC+SN+BAO+H0LiCOW data. These predictions are compatible with the $\Lambda$CDM cosmology, i.e., $\delta =0$.

\begin{figure*}
\begin{center}
\includegraphics[width=3.1in]{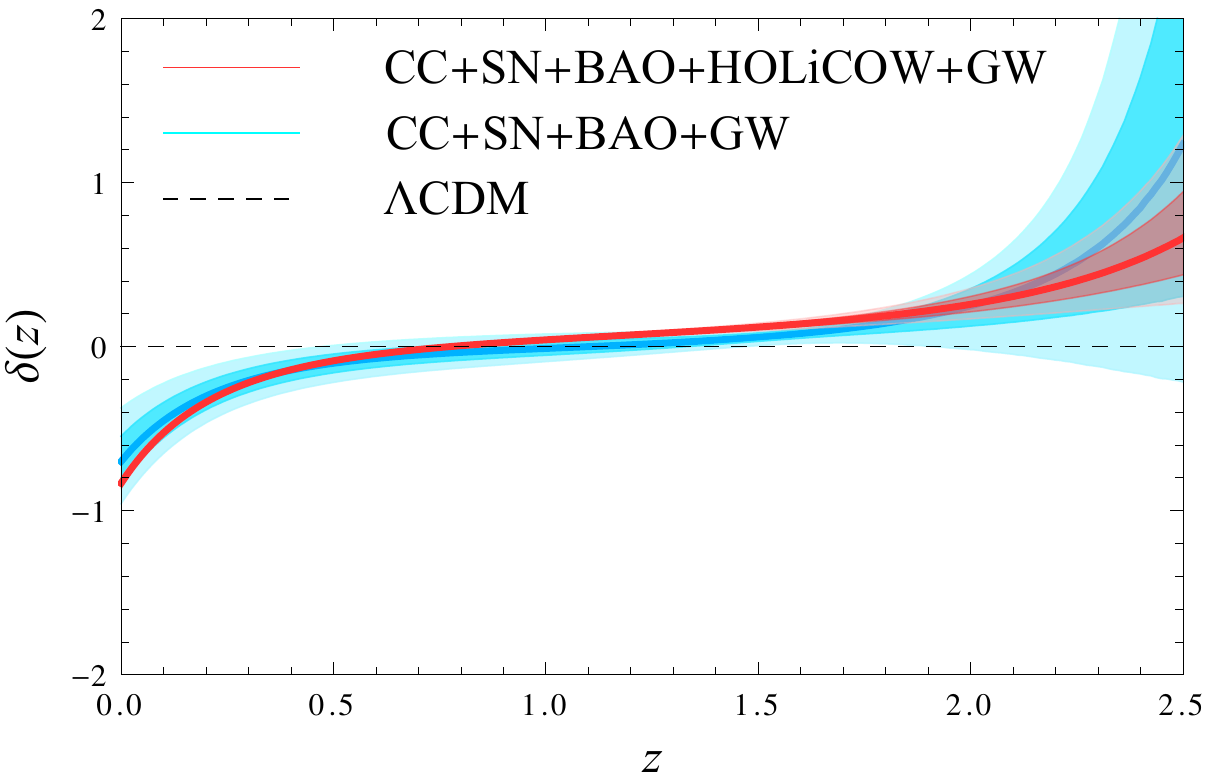} \,\,\,\,
\includegraphics[width=3.1in]{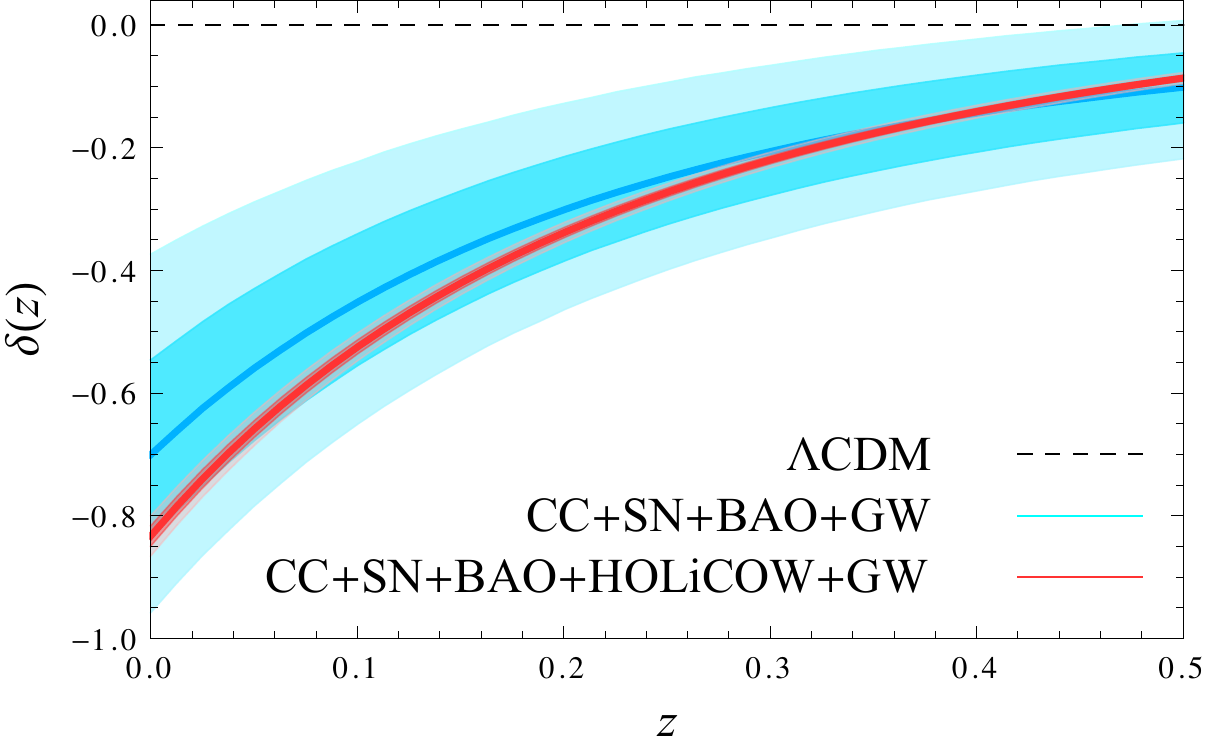}
\caption{Left-hand panel: Reconstructed coupling function $\delta (z)$ at $1\sigma$ and $2\sigma$ CL from CC+SN+BAO+GW (Blue) and CC+SN+BAO+H0LiCOW+GW (Red) data, in the interacting vacuum energy scenario. Right-hand panel: The same as in left-hand panel, but restricted to the range $z \in [0, 0.5]$. The dashed black curve corresponds to the canonical $\Lambda$CDM prediction and the solid curves are for the GP mean.}
\label{results_mock_data}
\end{center}
\end{figure*}

\begin{figure*}
\begin{center}
\includegraphics[width=3.1in]{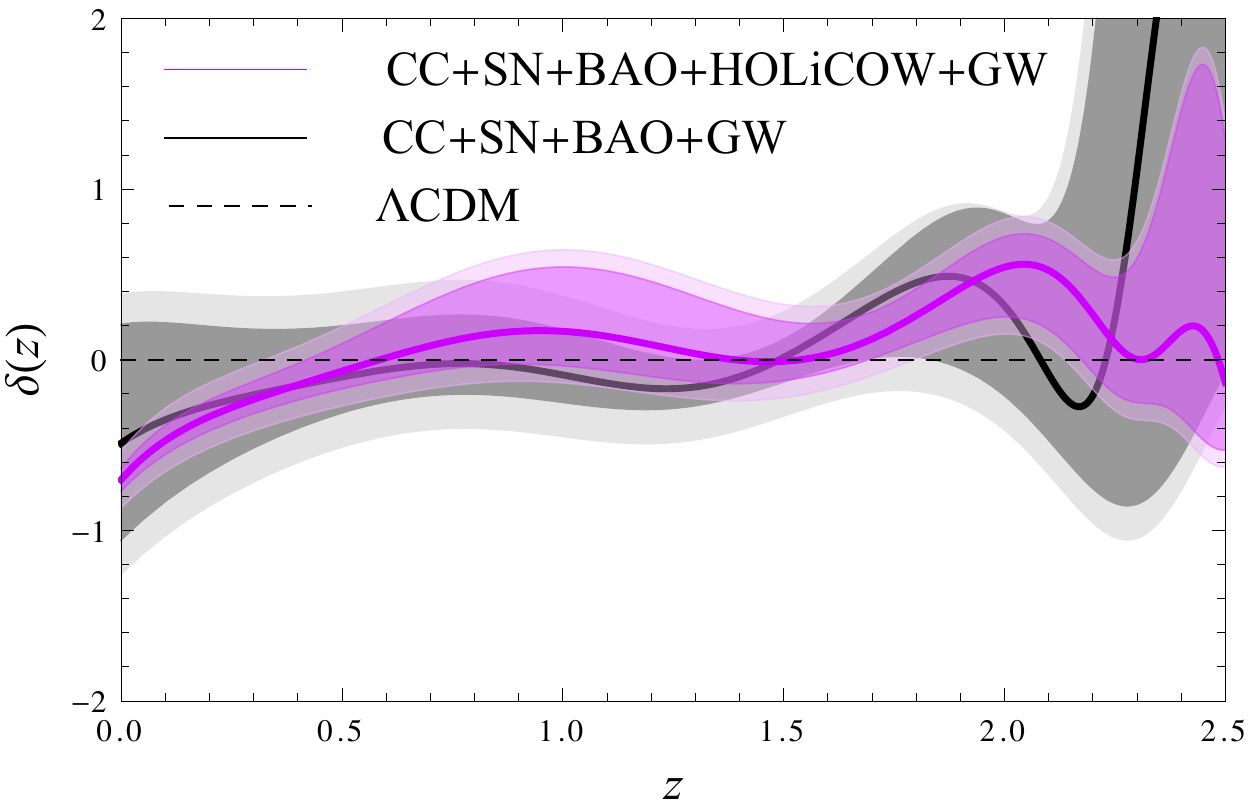} \,\,\,\,
\includegraphics[width=3.1in]{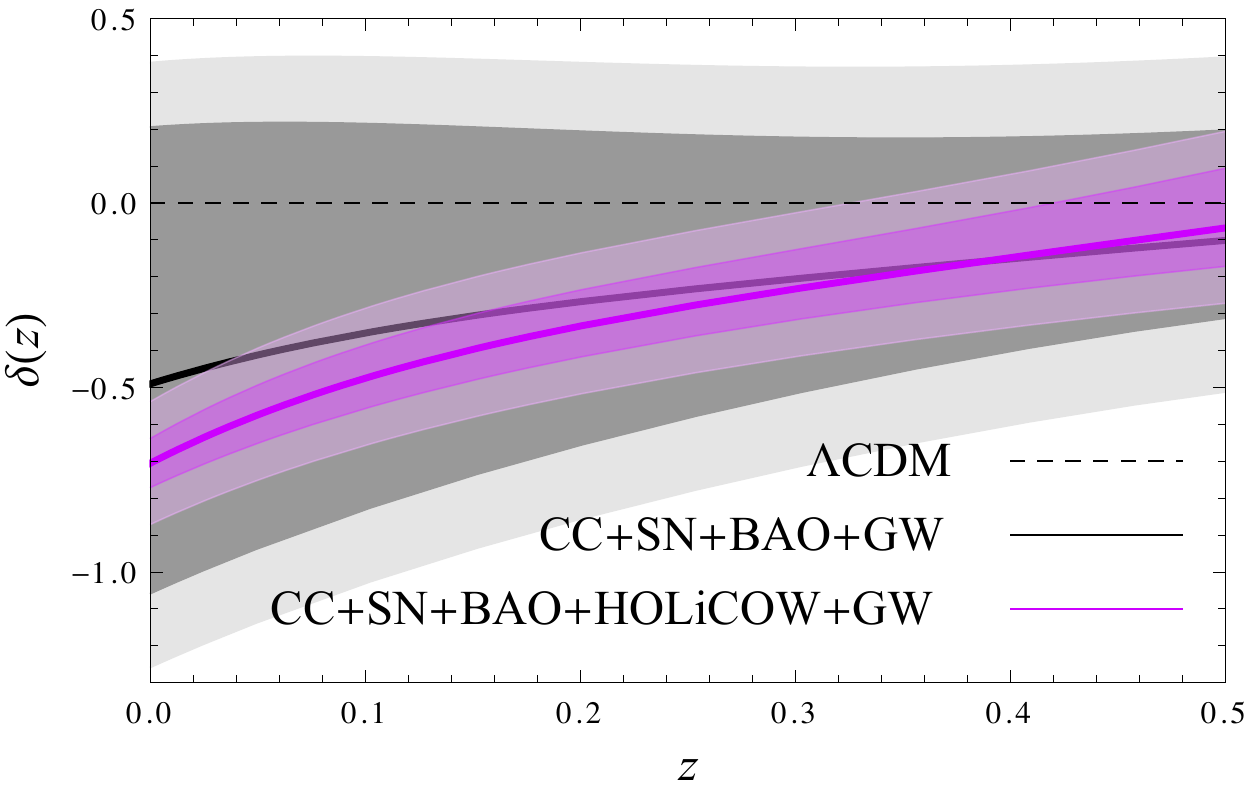}
\caption{Left-hand panel: Reconstructed coupling function $\delta (z)$ at $1\sigma$ and $2\sigma$ CL in the general interacting scenario of dark sectors' from CC+SN+BAO+GW (Black) and CC+SN+BAO+H0LiCOW+GW (Purple) data. Right-hand panel: The same as in left-hand panel, but restricted to the range $z \in [0, 0.5]$. The dashed black curve corresponds to the canonical $\Lambda$CDM prediction and the solid curves stand for the GP mean.}
\label{results_mock_data_w}
\end{center}
\end{figure*}

\section{Forecast from gravitational wave standard sirens}
\label{sec-gw}

To impose  more robust and accurate constraints on the $\delta (z)$ function, we  optimize the covariance function using mock gravitational waves (GW)  data generated by assuming the $\Lambda$CDM model as a fiducial one. As argued in \cite{Seikel2013}, for non-parametric regression method such as GP, we aim to generate confidence limits such that the true function is trapped appropriately. But it can be problematic when evaluating functions as $w$ and $\delta (z)$ because they are quantities that depend on the second and third order derivatives of the cosmological observable. We can avoid  the problem in identifying an appropriate covariance function which can reproduce expected models accurately by adding simulated data. To accomplish this object, we create a standard sirens mock catalogue, the gravitational wave analogue of the astronomical standard candles, which can provide powerful information about the dynamics of the universe. For a given GW strain signal, $h(t) = A(t) \cos [\Phi(t)]$, the stationary-phase approximation can be used for the orbital phase of inspiraling binary system to obtain its Fourier transform $\tilde{h}(f)$. For a coalescing binary system of masses $m_1$ and $m_2$, 

\begin{equation}
\label{waveform}
\tilde{h}(f) = Q \mathcal{A} f^{-7/6} e^{i\Phi(f)}.
\end{equation}

Here $\mathcal{A} \propto 1/d_L$ is the luminosity distance to the merger's redshift, and $\Phi(f)$ is the binary system's inspiral phase. For more details on the post-Newtonian coefficients and waveforms, one may  refer to \cite{Agostino_Nunes2019} and Appendix A therein. After defining the GW signal, for a high enough signal-to-noise ratio (SNR), one may obtain upper bounds on the free parameters of the GW signal $\tilde{h}(f)$ by using the Fisher information analysis. Estimating $d_L(z)$ from GW standard sirens mock data is well established approach, see \cite{Agostino_Nunes2019} and references therein. In what follows, we briefly describe our methodology that is used to generate the standard sirens mock catalogue.

In order to generate the mock standard siren catalogue, we consider the ET power spectral density noises. The ET is a third-generation ground detector, and  covers frequencies in the range $1-10^4$ Hz. The ET is sensitive to signal amplitude, which is expected to be ten times larger than the current advanced ground-based detectors. The ET conceptual design study predicts BNS detection of an order of $10^3-10^7$  per year. However, only a small fraction ($\sim 10^{-3}$) of them is expected to be accompanied by a short $\gamma$-ray burst observation. Assuming a detection rate of $\mathcal{O}(10^5)$, the events with short $\gamma$-ray bursts will be $\mathcal{O}(10^2)$ per year. 
In our simulations, 1000 BNS mock GW standard sirens merger events up to $z = 2$ are considered. In the mock catalogue, we have used  the input values $H_0 = 67.4$ $km$ $s^{-1}$  $Mpc^{-1}$ and $\Omega_{m0} = 0.31$  for the Hubble constant and matter density parameter, respectively, in agreement with the most recent Planck CMB data (within the $\Lambda$CDM paradigm) \cite{Aghanim:2018eyx}. We have estimated the measurement error on the luminosity distance for each event using the Fisher matrix analysis on the waveforms (see ref.~\cite{Agostino_Nunes2019} for details). We have calculated the SNR of each event, and confirmed that it is a GW detection provided SNR $> 8$. In what follows, we describe the evolution of the interaction function with the inclusion of the mock GW standard sirens with the standard cosmological probes. 

In Fig.~\ref{results_mock_data}, we show the  reconstructed interaction function $\delta(z)$ from CC+SN+BAO+GW and CC+SN+BAO+H0LiCOW+GW data combinations for the simple interaction scenario with $w = -1$. When evaluated at the present moment, we find $\delta(z=0) = -0.70 \pm 0.14$ at $1\sigma$ CL for CC+SN+BAO, and $\delta(z=0) = -0.833 \pm 0.016$ at $1\sigma$ CL for CC+SN+BAO+H0LiCOW under the interacting vacuum-energy assumption. Analysing the behavior of the $\delta$ function in the range $z \in [0, 2.5]$, we find an evidence for a sign transition in $\delta$ that quantifies the interaction between the dark components. We clearly notice a preference for $\delta < 0$ at late times. In Fig.~\ref{results_mock_data_w}, we show the reconstructed  interaction function $\delta(z)$ from CC+SN+BAO+GW and CC+SN+BAO+H0LiCOW+GW data combinations, under the general assumption where $w(z)$ is a free function of $z$. In this case, we find $\delta(z=0) = -0.49 \pm 0.69$ at $1\sigma$ CL from CC+SN+BAO+GW and $\delta(z=0) = -0.705 \pm 0.066$ at $1\sigma$ CL from CC+SN+BAO+H0LiCOW+GW. In our analysis, even including GW mock data in CC+SN+BAO, we note that $\delta$ is compatible with $\Lambda$CDM. On the other hand, from CC+SN+BAO+H0LiCOW+GW data,  we can notice a prediction for $\delta < 0$ at late times.

It is important to note that the presence of GWs mock catalogue (assuming a fiducial $\Lambda$CDM model) is used for the purpose of optimizing the covariance function, as argued previously. The results summarized in Fig.~\ref{results_mock_data_w} are the most realistic, being the ones for the most general case.

\section{Conclusions}

\label{sec-conclu}

In this work, we have presented some generalized aspects of dark sectors' interaction that may be of interest to the community: (i) We have investigated the case where DE can assume a dynamical character through its equation of state along with the simplest vacuum-energy case $w =-1$. (ii) We have studied  joint analyses of the dark sector interaction with several geometrical probes following a minimally model-dependent way of GP. (iii) We have optimized the covariance function using mock GW standard sirens  to better reconstruct the function $\delta(z)$. In short, all these pieces of investigation have led to more general and robust results which could be helpful in order to have a deeper understanding of the physics of the dark sector.  Our observations are as follows: We find, for both interacting vacuum and general scenario, that, $\delta (z)$ exhibits a transient nature according to the analyses from CC+SN+BAO and CC+SN+BAO+H0LiCOW data,  and at very late time, $\delta (z)$ enters into the negative region (see Figs. \ref{IVCDM_results01} and \ref{Geral_results}). This conclusion remains unaltered when we include the 1000 mock GW standard sirens to the above two combined data sets (see Figs. \ref{results_mock_data} and \ref{results_mock_data_w}). However, the indication of a late interaction is strongly pronounced in the context of the interacting vacuum scenario where we find that for the CC+SN+BAO data, $\delta (z = 0) \neq 0$ at more than $1\sigma$ CL, but  $\delta (z = 0) \neq 0$ remains valid at more than $3\sigma$ CL for CC+SN+BAO+H0LiCOW data.  This is an interesting result in this work because the transfer of energy among the dark sector components, which has been observed in some phenomenological models,  is not ruled out in light of the model-independent analysis, and also for the combined analyses that we have performed during the reconstruction.  However, concerning the general interacting picture, we see that $\delta (z =0) =0$ is compatible within $1\sigma$ for both CC+SN+BAO and CC+SN+BAO+H0LiCOW data.
When the GW standard sirens enter into the analysis, we find that for the interacting vacuum scenario, again $\delta (z =0) \neq 0$ at several standard deviations, for both CC+SN+BAO+GW and CC+SN+BAO+H0LiCOW+GW data. Whilst for the general scenario, even if for CC+SN+BAO+GW data, $\delta (z= 0)=0$ is compatible within $1\sigma$ but for  CC+SN+BAO+H0LiCOW+GW data, we find a strong preference of an interaction at several standard deviations. Summarizing the results,  we find that the model-independent analyses indicate for a possible interaction in the dark sector which is strongly preferred for the scenario with $w =-1$. Based on the findings of this study, we believe that it will be worthwhile to investigate, in future communications, various statistical techniques for reconstructing the function $\delta (z)$, such as the use of neural networks, principal component analysis, and others, which may provide statistical improvements over the standard GP method that we use in the study of cosmological parameters. \\

\section*{Acknowledgements}

\noindent 
The authors thank the referee for some useful comments that improved the manuscript. 
SK gratefully acknowledges the support from Science and Engineering Research Board (SERB), Govt. of India (File No. CRG/2021/004658). RCN would like to thank the agency FAPESP for financial support under the project No. 2018/18036-5.  SP acknowledges the Mathematical Research Impact-Centric Support Scheme (File No. MTR/2018/000940) of SERB, Govt. of India. 

%%%%%%%%%%%%%%%%%%%%%%%%%%%%%%%%%%%%%%%%%%%%%%%%%%
\section*{Data Availability}

The observational data used in this article will be shared on reasonable request to the corresponding author.

%%%%%%%%%%%%%%%%%%%% REFERENCES %%%%%%%%%%%%%%%%%%

% The best way to enter references is to use BibTeX:

%\bibliographystyle{mnras}
%\bibliography{example} % if your bibtex file is called example.bib

\begin{thebibliography}{99}

\bibitem[\protect\citeauthoryear{Akarsu et al.}{2021}]{2021PhRvD.104l3512A} Akarsu {\"O}., Kumar S., {\"O}z{\"u}lker E., Vazquez J.~A., 2021, PhRvD, 104, 123512. doi:10.1103/PhysRevD.104.123512

\bibitem[\protect\citeauthoryear{Alam et al.}{2017}]{Alam17} Alam S., Ata M., Bailey S., Beutler F., Bizyaev D., Blazek J.~A., Bolton A.~S., et al., 2017, MNRAS, 470, 2617. doi:10.1093/mnras/stx721

\bibitem[\protect\citeauthoryear{An, Feng, \& Wang}{2018}]{An:2017crg} An R., Feng C., Wang B., 2018, JCAP, 2018, 038. doi:10.1088/1475-7516/2018/02/038

\bibitem[\protect\citeauthoryear{Anchordoqui et al.}{2021}]{2021JHEAp..32...28A} Anchordoqui L.~A., Di Valentino E., Pan S., Yang W., 2021, JHEAp, 32, 28. doi:10.1016/j.jheap.2021.08.001

\bibitem[\protect\citeauthoryear{Avila et al.}{2022}]{2022MNRAS.509.2994A} Avila F., Bernui A., Nunes R.~C., de Carvalho E., Novaes C.~P., 2022, MNRAS, 509, 2994. doi:10.1093/mnras/stab3122

%\bibitem[\protect\citeauthoryear{Avila et al.}{2022}]{2022arXiv220107829A} Avila F., Bernui A., Bonilla A., Nunes R.~C., 2022, arXiv, arXiv:2201.07829

\bibitem[\protect\citeauthoryear{Avila et al}{2022}]{Avila:2022xad}
Avila F., Bernui A., Bonilla A., Nunes R. C., [arXiv:2201.07829 [astro-ph.CO]].

\bibitem[\protect\citeauthoryear{Bahamonde et al}{2021}]{Bahamonde:2021gfp}
S.~Bahamonde, K.~F.~Dialektopoulos, C.~Escamilla-Rivera, G.~Farrugia, V.~Gakis, M.~Hendry, M.~Hohmann, J.~L.~Said, J.~Mifsud and E.~Di Valentino, [arXiv:2106.13793 [gr-qc]].

%\bibitem[\protect\citeauthoryear{Bahamonde et al.}{2021}]{2021arXiv210613793B} Bahamonde S., Dialektopoulos K.~F., Escamilla-Rivera C., Farrugia G., Gakis V., Hendry M., Hohmann M., et al., 2021, arXiv, arXiv:2106.13793

%\bibitem[\protect\citeauthoryear{Bamba et al.}{2012}]{2012Ap&SS.342..155B} Bamba K., Capozziello S., Nojiri S., Odintsov S.~D., 2012, Ap\&SS, 342, 155. doi:10.1007/s10509-012-1181-8

\bibitem[\protect\citeauthoryear{Bamba et al}{2012}]{Bamba:2012cp}
K.~Bamba, S.~Capozziello, S.~Nojiri and S.~D.~Odintsov, Astrophys. Space Sci. \textbf{342}, 155-228 (2012), doi:10.1007/s10509-012-1181-8, [arXiv:1205.3421 [gr-qc]].

\bibitem[\protect\citeauthoryear{Bengaly}{2022}]{Bengaly:2021wgc}
Bengaly C., doi:10.1016/j.dark.2022.100966, [arXiv:2111.06869 [gr-qc]].

\bibitem[\protect\citeauthoryear{Bengaly, Clarkson, \& Maartens}{2020}]{GP_05} Bengaly C.~A.~P., Clarkson C., Maartens R., 2020, JCAP, 2020, 053. doi:10.1088/1475-7516/2020/05/053

%\bibitem[\protect\citeauthoryear{Bernardo}{2021}]{2021PhRvD.104b4070B} Bernardo R.~C., 2021, PhRvD, 104, 024070. doi:10.1103/PhysRevD.104.024070

\bibitem[\protect\citeauthoryear{Bernardo \& Levi Said}{2021}]{2021JCAP...09..014B} Bernardo R.~C., Levi Said J., 2021, JCAP, 2021, 014. doi:10.1088/1475-7516/2021/09/014


\bibitem[\protect\citeauthoryear{Bernardo et al.}{2021}]{Bernardo:2021cxi}
Bernardo R.~C.~, Grand\'on D., Said J.~L., C\'ardenas V.~H., [arXiv:2111.08289 [astro-ph.CO]].

\bibitem[\protect\citeauthoryear{Birrer et al.}{2019}]{Birrer2019} Birrer S., Treu T., Rusu C.~E., Bonvin V., Fassnacht C.~D., Chan J.~H.~H., Agnello A., et al., 2019, MNRAS, 484, 4726. doi:10.1093/mnras/stz200

\bibitem[\protect\citeauthoryear{B{\"o}hmer, Tamanini, \& Wright}{2015}]{Boehmer:2015kta} B{\"o}hmer C.~G., Tamanini N., Wright M., 2015, PhRvD, 91, 123002. doi:10.1103/PhysRevD.91.123002

\bibitem[\protect\citeauthoryear{Bolotin et al.}{2015}]{Bolotin:2013jpa} Bolotin Y.~L., Kostenko A., Lemets O.~A., Yerokhin D.~A., 2015, IJMPD, 24, 1530007. doi:10.1142/S0218271815300074

%\bibitem[\protect\citeauthoryear{Bonilla Rivera \& Castillo Hernandez}{2016}]{2016arXiv160100183B} Bonilla Rivera A., Castillo Hernandez J.~E., 2016, arXiv, arXiv:1601.00183

\bibitem[\protect\citeauthoryear{Bonilla Rivera \& Garc{\'\i}a Farieta}{2016}]{Bonilla_2} Bonilla Rivera A., Garc{\'\i}a Farieta J., 2016, arXiv, arXiv:1605.01984

\bibitem[\protect\citeauthoryear{Bonilla \& Castillo}{2018}]{Bonilla_1} Bonilla A., Castillo J., 2018, Univ, 4, 21. doi:10.3390/universe4010021

%\bibitem[\protect\citeauthoryear{Bonilla et al.}{2020}]{2020JCAP...03..015B} Bonilla A., D'Agostino R., Nunes R.~C., de Araujo J.~C.~N., 2020, JCAP, 2020, 015. doi:10.1088/1475-7516/2020/03/015

\bibitem[\protect\citeauthoryear{Bonilla, Kumar, \& Nunes}{2021}]{Bonilla:2020wbn} Bonilla A., Kumar S., Nunes R.~C., 2021, EPJC, 81, 127. doi:10.1140/epjc/s10052-021-08925-z

%\bibitem[\protect\citeauthoryear{Bora, Holanda, \& Desai}{2021}]{2021EPJC...81..596B} Bora K., Holanda R.~F.~L., Desai S., 2021, EPJC, 81, 596. doi:10.1140/epjc/s10052-021-09421-0

%\bibitem[\protect\citeauthoryear{Bora et al.}{2021}]{2021arXiv210615805B} Bora K., Holanda R.~F.~L., Desai S., Pereira S.~H., 2021, arXiv, arXiv:2106.15805

\bibitem[\protect\citeauthoryear{Briffa et al.}{2020}]{2020CQGra..38e5007B} Briffa R., Capozziello S., Said J.~L., Mifsud J., Saridakis E.~N., 2020, CQGra, 38, 055007. doi:10.1088/1361-6382/abd4f5

%\bibitem[\protect\citeauthoryear{Cai et al.}{2010}]{2010PhR...493....1C} Cai Y.-F., Saridakis E.~N., Setare M.~R., Xia J.-Q., 2010, PhR, 493, 1. doi:10.1016/j.physrep.2010.04.001

\bibitem[\protect\citeauthoryear{Cai et al}{2010}]{Cai:2009zp}
Y.~F.~Cai, E.~N.~Saridakis, M.~R.~Setare and J.~Q.~Xia, Phys. Rept. \textbf{493}, 1-60 (2010), doi:10.1016/j.physrep.2010.04.001, [arXiv:0909.2776 [hep-th]].

\bibitem[\protect\citeauthoryear{Cai et al}{2016}]{Cai:2015emx}
Y.~F.~Cai, S.~Capozziello, M.~De Laurentis and E.~N.~Saridakis, Rept. Prog. Phys. \textbf{79}, no.10, 106901 (2016), doi:10.1088/0034-4885/79/10/106901, [arXiv:1511.07586 [gr-qc]].

%\bibitem[\protect\citeauthoryear{Cai et al.}{2016}]{2016RPPh...79j6901C} Cai Y.-F., Capozziello S., De Laurentis M., Saridakis E.~N., 2016, RPPh, 79, 106901. doi:10.1088/0034-4885/79/10/106901

\bibitem[\protect\citeauthoryear{Cai, Tamanini, \& Yang}{2017}]{GP_IDE_01} Cai R.-G., Tamanini N., Yang T., 2017, JCAP, 2017, 031. doi:10.1088/1475-7516/2017/05/031

%\bibitem[\protect\citeauthoryear{Capozziello \& de Laurentis}{2011}]{2011PhR...509..167C} Capozziello S., de Laurentis M., 2011, PhR, 509, 167. doi:10.1016/j.physrep.2011.09.003

\bibitem[\protect\citeauthoryear{Capozziello and De Laurentis}{2011}]{Capozziello:2011et} S.~Capozziello and M.~De Laurentis, Phys. Rept. \textbf{509}, 167-321 (2011), doi:10.1016/j.physrep.2011.09.003, [arXiv:1108.6266 [gr-qc]].

\bibitem[\protect\citeauthoryear{Clifton et al}{2012}]{Clifton:2011jh}
T.~Clifton, P.~G.~Ferreira, A.~Padilla and C.~Skordis,Phys. Rept. \textbf{513}, 1-189 (2012), doi:10.1016/j.physrep.2012.01.001, [arXiv:1106.2476 [astro-ph.CO]].

%\bibitem[\protect\citeauthoryear{Clifton et al.}{2012}]{2012PhR...513....1C} Clifton T., Ferreira P.~G., Padilla A., Skordis C., 2012, PhR, 513, 1. doi:10.1016/j.physrep.2012.01.001

\bibitem[\protect\citeauthoryear{Colg{\'a}in \& Sheikh-Jabbari}{2021}]{GP_09} Colg{\'a}in E. {\'O}., Sheikh-Jabbari M.~M., 2021, arXiv, arXiv:2101.08565

\bibitem[\protect\citeauthoryear{Copeland, Sami \& Tsujikawa}{2006}]{Copeland:2006wr}
E.~J.~Copeland, M.~Sami and S.~Tsujikawa, Int. J. Mod. Phys. D \textbf{15}, 1753-1936 (2006)

%\bibitem[\protect\citeauthoryear{Colg{\'a}in, Sheikh-Jabbari, \& Yin}{2021}]{2021arXiv210401930C} Colg{\'a}in E. {\'O}., Sheikh-Jabbari M.~M., Yin L., 2021, arXiv, arXiv:2104.01930

%\bibitem[\protect\citeauthoryear{Colg{\'a}in \& Sheikh-Jabbari}{2021}]{2021arXiv210108565C} Colg{\'a}in E. {\'O}., Sheikh-Jabbari M.~M., 2021, arXiv, arXiv:2101.08565

\bibitem[\protect\citeauthoryear{D'Agostino \& Nunes}{2019}]{Agostino_Nunes2019} D'Agostino R., Nunes R.~C., 2019, PhRvD, 100, 044041. doi:10.1103/PhysRevD.100.044041

\bibitem[\protect\citeauthoryear{D'Amico, Hamill, \& Kaloper}{2016}]{Amico:2016qft} D'Amico G., Hamill T., Kaloper N., 2016, PhRvD, 94, 103526. doi:10.1103/PhysRevD.94.103526

\bibitem[\protect\citeauthoryear{de Araujo et al.}{2021}]{2021PhRvD.104j4057D} de Araujo J.~C.~N., De Felice A., Kumar S., Nunes R.~C., 2021, PhRvD, 104, 104057. doi:10.1103/PhysRevD.104.104057

%\bibitem[\protect\citeauthoryear{De Felice \& Tsujikawa}{2010}]{2010LRR....13....3D} De Felice A., Tsujikawa S., 2010, LRR, 13, 3. doi:10.12942/lrr-2010-3

\bibitem[\protect\citeauthoryear{De Felice S.~Tsujikawa}{2010}]{DeFelice:2010aj}
A.~De Felice and S.~Tsujikawa, Living Rev. Rel. \textbf{13}, 3 (2010), doi:10.12942/lrr-2010-3, [arXiv:1002.4928 [gr-qc]].

\bibitem[\protect\citeauthoryear{del Campo, Herrera, \& Pav\'on}{2009}]{delCampo:2008jx} del Campo S., Herrera R., Pav{\'o}n D., 2009, JCAP, 2009, 020. doi:10.1088/1475-7516/2009/01/020

\bibitem[\protect\citeauthoryear{Di Valentino, Melchiorri, \& Mena}{2017}]{DiValentino:2017iww} Di Valentino E., Melchiorri A., Mena O., 2017, PhRvD, 96, 043503. doi:10.1103/PhysRevD.96.043503

\bibitem[\protect\citeauthoryear{Di Valentino et al.}{2020}]{DiValentino:2019ffd} Di Valentino E., Melchiorri A., Mena O., Vagnozzi S., 2020, PDU, 30, 100666. doi:10.1016/j.dark.2020.100666

\bibitem[\protect\citeauthoryear{Di Valentino et al.}{2021}]{S8_tension} Di Valentino E., Anchordoqui L.~A., Akarsu {\"O}., Ali-Haimoud Y., Amendola L., Arendse N., Asgari M., et al., 2021, APh, 131, 102604. doi:10.1016/j.astropartphys.2021.102604

\bibitem[\protect\citeauthoryear{Di Valentino et al.}{2021}]{2021CQGra..38o3001D} Di Valentino E., Mena O., Pan S., Visinelli L., Yang W., Melchiorri A., Mota D.~F., et al., 2021, CQGra, 38, 153001. doi:10.1088/1361-6382/ac086d

%\bibitem[\protect\citeauthoryear{Di Valentino et al}{2021}]{DiValentino:2021izs} E.~Di Valentino, O.~Mena, S.~Pan, L.~Visinelli, W.~Yang, A.~Melchiorri, D.~F.~Mota, A.~G.~Riess and J.~Silk, Class. Quant. Grav. \textbf{38}, no.15, 153001 (2021)

\bibitem[\protect\citeauthoryear{Di Valentino}{2021}]{DiValentino:2021pow} Di Valentino E., 2021, Springer International Publishing, 483--505, doi:10.1007/978-3-030-83715-0\_32

%\bibitem[\protect\citeauthoryear{Dialektopoulos et al.}{2021}]{Dialektopoulos:2021wde} Dialektopoulos K., Said J.~L., Mifsud J., Sultana J., Adami K.~Z., [arXiv:2111.11462 [astro-ph.CO]].

\bibitem[\protect\citeauthoryear{Dialektopoulos et al.}{2022}]{Dialektopoulos:2021wde} Dialektopoulos K., Said J.~L., Mifsud J., Sultana J., Zarb Adami K., 2022, JCAP, 2022, 023. doi:10.1088/1475-7516/2022/02/023

\bibitem[\protect\citeauthoryear{du Mas des Bourboux et al.}{2020}]{du_Mas20} du Mas des Bourboux H., Rich J., Font-Ribera A., de Sainte Agathe V., Farr J., Etourneau T., Le Goff J.-M., et al., 2020, ApJ, 901, 153. doi:10.3847/1538-4357/abb085

\bibitem[\protect\citeauthoryear{du Mas des Bourboux et al.}{2017}]{du_Mas17} du Mas des Bourboux H., Le Goff J.-M., Blomqvist M., Busca N.~G., Guy J., Rich J., Y{\`e}che C., et al., 2017, A\&A, 608, A130. doi:10.1051/0004-6361/201731731

\bibitem[\protect\citeauthoryear{Elizalde, Gluza, \& Khurshudyan}{2021}]{2021arXiv210401077E} Elizalde E., Gluza J., Khurshudyan M., 2021, arXiv, arXiv:2104.01077

\bibitem[\protect\citeauthoryear{Gleyzes et al.}{2015}]{Gleyzes:2015pma} Gleyzes J., Langlois D., Mancarella M., Vernizzi F., 2015, JCAP, 2015, 054. doi:10.1088/1475-7516/2015/08/054

\bibitem[\protect\citeauthoryear{G{\'o}mez-Valent \& Amendola}{2018}]{GP_08} G{\'o}mez-Valent A., Amendola L., 2018, JCAP, 2018, 051. doi:10.1088/1475-7516/2018/04/051

\bibitem[\protect\citeauthoryear{Haridasu et al.}{2018}]{Haridasu18} Haridasu B.~S., Lukovi{\'c} V.~V., Moresco M., Vittorio N., 2018, JCAP, 2018, 015. doi:10.1088/1475-7516/2018/10/015

\bibitem[\protect\citeauthoryear{Jesus et al.}{2020}]{Jesus2020} Jesus J.~F., Valentim R., Escobal A.~A., Pereira S.~H., 2020, JCAP, 2020, 053. doi:10.1088/1475-7516/2020/04/053

\bibitem[\protect\citeauthoryear{Kase \& Tsujikawa}{2020}]{Kase:2019hor} Kase R., Tsujikawa S., 2020, PhRvD, 101, 063511. doi:10.1103/PhysRevD.101.063511

\bibitem[\protect\citeauthoryear{Kumar \& Nunes}{2017}]{Kumar:2017dnp} Kumar S., Nunes R.~C., 2017, PhRvD, 96, 103511. doi:10.1103/PhysRevD.96.103511

\bibitem[\protect\citeauthoryear{Kumar, Nunes, \& Yadav}{2019}]{Kumar:2019wfs} Kumar S., Nunes R.~C., Yadav S.~K., 2019, EPJC, 79, 576. doi:10.1140/epjc/s10052-019-7087-7

\bibitem[\protect\citeauthoryear{Kumar}{2021}]{2021PDU....3300862K} Kumar S., 2021, PDU, 33, 100862. doi:10.1016/j.dark.2021.100862


\bibitem[\protect\citeauthoryear{Liao et al.}{2019}]{GP_07} Liao K., Shafieloo A., Keeley R.~E., Linder E.~V., 2019, ApJL, 886, L23. doi:10.3847/2041-8213/ab5308

\bibitem[\protect\citeauthoryear{Lorenz et al.}{2021}]{2021PhRvD.104l3518L} Lorenz C.~S., Funcke L., L{\"o}ffler M., Calabrese E., 2021, PhRvD, 104, 123518. doi:10.1103/PhysRevD.104.123518

\bibitem[\protect\citeauthoryear{Lucca \& Hooper}{2020}]{Lucca:2020zjb} Lucca M., Hooper D.~C., 2020, PhRvD, 102, 123502. doi:10.1103/PhysRevD.102.123502

\bibitem[\protect\citeauthoryear{Lucca}{2021}]{2021PDU....3400899L} Lucca M., 2021, PDU, 34, 100899. doi:10.1016/j.dark.2021.100899

\bibitem[\protect\citeauthoryear{Lucca}{2021}]{2021PhRvD.104h3510L} Lucca M., 2021, PhRvD, 104, 083510. doi:10.1103/PhysRevD.104.083510

\bibitem[\protect\citeauthoryear{Martinelli et al.}{2019}]{GP_IDE_04} Martinelli M., Hogg N.~B., Peirone S., Bruni M., Wands D., 2019, MNRAS, 488, 3423. doi:10.1093/mnras/stz1915

\bibitem[\protect\citeauthoryear{Mehrabi \& Rezaei}{2021}]{2021arXiv211014950M} Mehrabi A., Rezaei M., 2021, arXiv, arXiv:2110.14950

\bibitem[\protect\citeauthoryear{Mohseni Sadjadi \& Honardoost}{2007}]{Sadjadi:2006qb} Mohseni Sadjadi H., Honardoost M., 2007, PhLB, 647, 231. doi:10.1016/j.physletb.2007.02.016

%\bibitem[\protect\citeauthoryear{Moradpour et al.}{2017}]{2017PhRvD..96l3504M} Moradpour H., Bonilla A., Abreu E.~M.~C., Neto J.~A., 2017, PhRvD, 96, 123504. doi:10.1103/PhysRevD.96.123504

\bibitem[\protect\citeauthoryear{Moresco et al.}{2016}]{Moresco16} Moresco M., Pozzetti L., Cimatti A., Jimenez R., Maraston C., Verde L., Thomas D., et al., 2016, JCAP, 2016, 014. doi:10.1088/1475-7516/2016/05/014

\bibitem[\protect\citeauthoryear{Mukherjee \& Banerjee}{2017}]{GP_IDE_02} Mukherjee A., Banerjee N., 2017, CQGra, 34, 035016. doi:10.1088/1361-6382/aa54c8

%\bibitem[\protect\citeauthoryear{Nojiri, Odintsov, \& Oikonomou}{2017}]{2017PhR...692....1N} Nojiri S., Odintsov S.~D., Oikonomou V.~K., 2017, PhR, 692, 1. doi:10.1016/j.physrep.2017.06.001

\bibitem[\protect\citeauthoryear{Nojiri, Odintsov \& Oikonomou}{2017}]{Nojiri:2017ncd}
S.~Nojiri, S.~D.~Odintsov and V.~K.~Oikonomou, Phys. Rept. \textbf{692}, 1-104 (2017)
doi:10.1016/j.physrep.2017.06.001

%\bibitem[\protect\citeauthoryear{Nunes et al.}{2017}]{2017EPJC...77..230N} Nunes R.~C., Bonilla A., Pan S., Saridakis E.~N., 2017, EPJC, 77, 230. doi:10.1140/epjc/s10052-017-4798-5

\bibitem[\protect\citeauthoryear{Nunes et al.}{2020}]{GP_06} Nunes R.~C., Yadav S.~K., Jesus J.~F., Bernui A., 2020, MNRAS, 497, 2133. doi:10.1093/mnras/staa2036

%\bibitem[\protect\citeauthoryear{Starobinsky}{1998}]{w_01} Starobinsky A.~A., 1998, JETPL, 68, 757. doi:10.1134/1.567941

%\bibitem[\protect\citeauthoryear{Nakamura \& Chiba}{1999}]{w_02} Nakamura T., Chiba T., 1999, MNRAS, 306, 696. doi:10.1046/j.1365-8711.1999.02551.x

%\bibitem[\protect\citeauthoryear{Huterer \& Turner}{1999}]{w_03} Huterer D., Turner M.~S., 1999, PhRvD, 60, 081301. doi:10.1103/PhysRevD.60.081301

\bibitem[\protect\citeauthoryear{Pan \& Chakraborty}{2014}]{Pan:2014afa} Pan S., Chakraborty S., 2014, IJMPD, 23, 1450092. doi:10.1142/S0218271814500928

\bibitem[\protect\citeauthoryear{Pan et al.}{2019}]{Pan:2019jqh} Pan S., Yang W., Singha C., Saridakis E.~N., 2019, PhRvD, 100, 083539. doi:10.1103/PhysRevD.100.083539

\bibitem[\protect\citeauthoryear{Pan, Yang, \& Paliathanasis}{2020}]{Pan:2020bur} Pan S., Yang W., Paliathanasis A., 2020, MNRAS, 493, 3114. doi:10.1093/mnras/staa213

\bibitem[\protect\citeauthoryear{Pan, Sharov, \& Yang}{2020}]{Pan:2020zza} Pan S., Sharov G.~S., Yang W., 2020, PhRvD, 101, 103533. doi:10.1103/PhysRevD.101.103533

\bibitem[\protect\citeauthoryear{Pandey, Raveri, \& Jain}{2020}]{Pandey2020} Pandey S., Raveri M., Jain B., 2020, PhRvD, 102, 023505. doi:10.1103/PhysRevD.102.023505

\bibitem[\protect\citeauthoryear{Perivolaropoulos \& Skara}{2021}]{2021arXiv210505208P} Perivolaropoulos L., Skara F., 2021, arXiv, arXiv:2105.05208

\bibitem[\protect\citeauthoryear{Piedipalumbo, De Laurentis, \& Capozziello}{2020}]{2020PDU....2700444P} Piedipalumbo E., De Laurentis M., Capozziello S., 2020, PDU, 27, 100444. doi:10.1016/j.dark.2019.100444

\bibitem[\protect\citeauthoryear{Planck Collaboration et al.}{2020}]{Aghanim:2018eyx} Planck Collaboration, Aghanim N., Akrami Y., Ashdown M., Aumont J., Baccigalupi C., Ballardini M., et al., 2020, A\&A, 641, A6. doi:10.1051/0004-6361/201833910

%\bibitem[\protect\citeauthoryear{Pourhassan et al.}{2018}]{2018PDU....20...41P} Pourhassan B., Bonilla A., Faizal M., Abreu E.~M.~C., 2018, PDU, 20, 41. doi:10.1016/j.dark.2018.02.006

\bibitem[\protect\citeauthoryear{Pourtsidou \& Tram}{2016}]{Pourtsidou:2016ico} Pourtsidou A., Tram T., 2016, PhRvD, 94, 043518. doi:10.1103/PhysRevD.94.043518

\bibitem[\protect\citeauthoryear{Ren et al.}{2021}]{2021PDU....3200812R} Ren X., Wong T.~H.~T., Cai Y.-F., Saridakis E.~N., 2021, PDU, 32, 100812. doi:10.1016/j.dark.2021.100812

\bibitem[\protect\citeauthoryear{Renzi \& Silvestri}{2020}]{GP_10} Renzi F., Silvestri A., 2020, arXiv, arXiv:2011.10559

\bibitem[\protect\citeauthoryear{Renzi, Hogg, \& Giar{\`e}}{2021}]{2021arXiv211205701R} Renzi F., Hogg N.~B., Giar{\`e} W., 2021, arXiv, arXiv:2112.05701

\bibitem[\protect\citeauthoryear{Reyes \& Escamilla-Rivera}{2021}]{2021JCAP...07..048R} Reyes M., Escamilla-Rivera C., 2021, JCAP, 2021, 048. doi:10.1088/1475-7516/2021/07/048

\bibitem[\protect\citeauthoryear{Riess et al.}{2018}]{Riess18} Riess A.~G., Rodney S.~A., Scolnic D.~M., Shafer D.~L., Strolger L.-G., Ferguson H.~C., Postman M., et al., 2018, ApJ, 853, 126. doi:10.3847/1538-4357/aaa5a9

\bibitem[\protect\citeauthoryear{Riess et al.}{2019}]{Riess:2019cxk} Riess A.~G., Casertano S., Yuan W., Macri L.~M., Scolnic D., 2019, ApJ, 876, 85. doi:10.3847/1538-4357/ab1422

\bibitem[\protect\citeauthoryear{Riess et al.}{2021}]{2021arXiv211204510R} Riess A.~G., Yuan W., Macri L.~M., Scolnic D., Brout D., Casertano S., Jones D.~O., et al., 2021, arXiv, arXiv:2112.04510

%\bibitem[\protect\citeauthoryear{Sch{\"o}neberg et al.}{2021}]{2021arXiv210710291S} Sch{\"o}neberg N., Abell{\'a}n G.~F., P{\'e}rez S{\'a}nchez A., Witte S.~J., Poulin V., Lesgourgues J., 2021, arXiv, arXiv:2107.10291

\bibitem[\protect\citeauthoryear{Sch\"oneberg et al}{2021}]{Schoneberg:2021qvd}
N.~Sch\"oneberg, G.~Franco Abell\'an, A.~P\'erez S\'anchez, S.~J.~Witte, V.~Poulin and J.~Lesgourgues, [arXiv:2107.10291 [astro-ph.CO]].

\bibitem[\protect\citeauthoryear{Scolnic et al.}{2018}]{Scolnic18} Scolnic D.~M., Jones D.~O., Rest A., Pan Y.~C., Chornock R., Foley R.~J., Huber M.~E., et al., 2018, ApJ, 859, 101. doi:10.3847/1538-4357/aab9bb

\bibitem[\protect\citeauthoryear{Seikel \& Clarkson}{2013}]{Seikel2013} Seikel M., Clarkson C., 2013, arXiv, arXiv:1311.6678

\bibitem[\protect\citeauthoryear{Seikel, Clarkson, \& Smith}{2012}]{GP_01} Seikel M., Clarkson C., Smith M., 2012, JCAP, 2012, 036. doi:10.1088/1475-7516/2012/06/036

\bibitem[\protect\citeauthoryear{Shafieloo, Kim, \& Linder}{2012}]{GP_02} Shafieloo A., Kim A.~G., Linder E.~V., 2012, PhRvD, 85, 123530. doi:10.1103/PhysRevD.85.123530

%\bibitem[\protect\citeauthoryear{Sotiriou \& Faraoni}{2010}]{2010RvMP...82..451S} Sotiriou T.~P., Faraoni V., 2010, RvMP, 82, 451. doi:10.1103/RevModPhys.82.451

\bibitem[\protect\citeauthoryear{Sotiriou and Faraoni}{2010}]{Sotiriou:2008rp}
T.~P.~Sotiriou and V.~Faraoni, Rev. Mod. Phys. \textbf{82}, 451-497 (2010), doi:10.1103/RevModPhys.82.451, [arXiv:0805.1726 [gr-qc]].

\bibitem[\protect\citeauthoryear{Sun, Jiao, \& Zhang}{2021}]{2021ApJ...915..123S} Sun W., Jiao K., Zhang T.-J., 2021, ApJ, 915, 123. doi:10.3847/1538-4357/ac05b8

\bibitem[\protect\citeauthoryear{Suzuki et al.}{2012}]{Suzuki:2011hu} Suzuki N., Rubin D., Lidman C., Aldering G., Amanullah R., Barbary K., Barrientos L.~F., et al., 2012, ApJ, 746, 85. doi:10.1088/0004-637X/746/1/85

\bibitem[\protect\citeauthoryear{Theodoropoulos \& Perivolaropoulos}{2021}]{2021Univ....7..300T} Theodoropoulos A., Perivolaropoulos L., 2021, Univ, 7, 300. doi:10.3390/universe7080300

%\bibitem[\protect\citeauthoryear{Vagnozzi, Loeb, \& Moresco}{2021}]{2021ApJ...908...84V} Vagnozzi S., Loeb A., Moresco M., 2021, ApJ, 908, 84. doi:10.3847/1538-4357/abd4df

\bibitem[\protect\citeauthoryear{Vanrietvelde et al.}{2018}]{2018arXiv180900556V} Vanrietvelde A., Hoehn P.~A., Giacomini F., Castro-Ruiz E., 2018, arXiv, arXiv:1809.00556

\bibitem[\protect\citeauthoryear{Velten, vom Marttens, \& Zimdahl}{2014}]{Velten} Velten H.~E.~S., vom Marttens R.~F., Zimdahl W., 2014, EPJC, 74, 3160. doi:10.1140/epjc/s10052-014-3160-4

\bibitem[\protect\citeauthoryear{von Marttens et al.}{2019}]{GP_IDE_03} von Marttens R., Marra V., Casarini L., Gonzalez J.~E., Alcaniz J., 2019, PhRvD, 99, 043521. doi:10.1103/PhysRevD.99.043521

\bibitem[\protect\citeauthoryear{von Marttens et al.}{2021}]{GP_IDE_07} von Marttens R., Gonzalez J.~E., Alcaniz J., Marra V., Casarini L., 2021, PhRvD, 104, 043515. doi:10.1103/PhysRevD.104.043515

\bibitem[\protect\citeauthoryear{Wang \& Meng}{2017}]{GP_04} Wang D., Meng X.-H., 2017, PhRvD, 95, 023508. doi:10.1103/PhysRevD.95.023508

\bibitem[\protect\citeauthoryear{Wang, Gong, \& Abdalla}{2005}]{Wang:2005jx} Wang B., Gong Y., Abdalla E., 2005, PhLB, 624, 141. doi:10.1016/j.physletb.2005.08.008

\bibitem[\protect\citeauthoryear{Wang et al.}{2015}]{Wang2015} Wang Y., Zhao G.-B., Wands D., Pogosian L., Crittenden R.~G., 2015, PhRvD, 92, 103005. doi:10.1103/PhysRevD.92.103005

\bibitem[\protect\citeauthoryear{Wang et al.}{2016}]{Wang:2016lxa} Wang B., Abdalla E., Atrio-Barandela F., Pav{\'o}n D., 2016, RPPh, 79, 096901. doi:10.1088/0034-4885/79/9/096901

\bibitem[\protect\citeauthoryear{Weinberg}{1998}]{Weinberg:1988cp}
Weinberg ~S., 1989, Rev. Mod. Phys. \textbf{61}. doi:10.1103/RevModPhys.61.1

\bibitem[\protect\citeauthoryear{Wong et al.}{2020}]{H0LiCOW} Wong K.~C., Suyu S.~H., Chen G.~C.-F., Rusu C.~E., Millon M., Sluse D., Bonvin V., et al., 2020, MNRAS, 498, 1420. doi:10.1093/mnras/stz3094

\bibitem[\protect\citeauthoryear{Yang, Guo, \& Cai}{2015}]{Yang2015} Yang T., Guo Z.-K., Cai R.-G., 2015, PhRvD, 91, 123533. doi:10.1103/PhysRevD.91.123533

\bibitem[\protect\citeauthoryear{Yang}{2020}]{GP_IDE_06} Yang T., 2020, PhRvD, 102, 083511. doi:10.1103/PhysRevD.102.083511

\bibitem[\protect\citeauthoryear{Yang et al.}{2018}]{Yang:2018uae} Yang W., Mukherjee A., Di Valentino E., Pan S., 2018, PhRvD, 98, 123527. doi:10.1103/PhysRevD.98.123527

\bibitem[\protect\citeauthoryear{Yang et al.}{2021}]{2021PhRvD.103h3520Y} Yang W., Pan S., Sal{\'o} L.~A., de Haro J., 2021, PhRvD, 103, 083520. doi:10.1103/PhysRevD.103.083520

\bibitem[\protect\citeauthoryear{Yang et al.}{2021}]{2021JCAP...10..008Y} Yang W., Pan S., Di Valentino E., Mena O., Melchiorri A., 2021, JCAP, 2021, 008. doi:10.1088/1475-7516/2021/10/008

\bibitem[\protect\citeauthoryear{Zhang \& Xia}{2016}]{GP_03} Zhang M.-J., Xia J.-Q., 2016, JCAP, 2016, 005. doi:10.1088/1475-7516/2016/12/005

\bibitem[\protect\citeauthoryear{Zhao et al.}{2019}]{Zhao19} Zhao G.-B., Wang Y., Saito S., Gil-Mar{\'\i}n H., Percival W.~J., Wang D., Chuang C.-H., et al., 2019, MNRAS, 482, 3497. doi:10.1093/mnras/sty2845

\bibitem[\protect\citeauthoryear{Zhou, Zhang, \& Li}{2019}]{GP_IDE_05} Zhou Z., Zhang T.~J., Li T.~P., 2019, arXiv, arXiv:1908.06254

\end{thebibliography}

% Alternatively you could enter them by hand, like this:
% This method is tedious and prone to error if you have lots of references
%\begin{thebibliography}{99}
%\bibitem[\protect\citeauthoryear{Author}{2012}]{Author2012}
%Author A.~N., 2013, Journal of Improbable Astronomy, 1, 1
%\bibitem[\protect\citeauthoryear{Others}{2013}]{Others2013}
%Others S., 2012, Journal of Interesting Stuff, 17, 198
%\end{thebibliography}

%%%%%%%%%%%%%%%%%%%%%%%%%%%%%%%%%%%%%%%%%%%%%%%%%%

%%%%%%%%%%%%%%%%% APPENDICES %%%%%%%%%%%%%%%%%%%%%

\appendix

\section{Derivation of the key equation}
\label{sec-appendix}

Here we show the derivation of eqn. (\ref{eqn:WqE}) which is the heart of this work. Now, subtracting  eq.(\ref{cont2})  from eq.(\ref{cont1}) one can obtain

\begin{equation}\label{new-eqn}
    \left[ \dot{\rho}_{\rm DM} - \dot{\rho}_{\rm DE} \right] + 3 H \left[ \rho_{\rm DM}  - \left(1 + w\right) \rho_{\rm DE} \right] = - 2Q(t).
\end{equation}
Now, using the Friedmann's equations $3 H^2 =  \rho_{\rm DM} + \rho_{\rm DE}$ and  $2 \dot{H} + 3 H^2 = - p_{\rm DE} = - w\rho_{\rm DE}$, one can express $\rho_{\rm DM}$ and $\rho_{\rm DE}$ respectively as 

\begin{equation}
\rho_{\rm DM} = 3 H^2 +\frac{2\dot{H}}{w} + \frac{3H^2}{w},\label{eqn-rhoDM-appendix}
\end{equation}
and
\begin{equation}
\rho_{\rm DE}  = - \frac{2\dot{H}}{w} - \frac{3H^2}{w}\label{eqn-rhoDE-appendix}
\end{equation}

Now differentiating (\ref{eqn-rhoDM-appendix})  and (\ref{eqn-rhoDE-appendix}) with respect to the cosmic time 
one can write down:

\begin{equation}\label{dot-rhoDM}
    \dot{\rho}_{\rm DM} =  6H\dot{H}  + 2 \frac{\ddot{H}}{w} - 2 \frac{\dot{H} \dot{w}}{w^2} + 6 \frac{H\dot{H}}{w} - 3 \frac{H^2\dot{w}}{w^2},
\end{equation}
and 
\begin{equation}\label{dot-rhoDE}
    \dot{\rho}_{\rm DE} = - 2 \frac{\ddot{H}}{w} +  2 \frac{\dot{H} \dot{w}}{w^2} - 6 \frac{H\dot{H}}{w} + 3 \frac{H^2\dot{w}}{w^2},
\end{equation}

Notice that, by changing $\frac{d}{dt} \rightarrow \frac{d}{dz}$ where $z$ denotes the redshift, one can express 
\begin{eqnarray}
\dot{w} = - \frac{w' H}{a},\;\; \quad \dot{H} = - (1+z) H H',
\end{eqnarray}
where prime denotes the derivative with respect to $z$. Consequently, one can express $\ddot{H}$ as 
\begin{equation}\label{ddot-H}
\ddot{H} = (1+z)^2 H H'^2 + (1+z)^2 H^2 H'' + (1+z) H^2 H'
\end{equation}

Now, inserting eqns. (\ref{dot-rhoDM}), (\ref{dot-rhoDE}) into eqn. (\ref{new-eqn}) and using the derivatives of cosmological parameters with respect to $z$, and doing some simple algebraic calculations, one arrives at 
\begin{eqnarray}
- w Q &=& 2 \Big(H H'^2 + H^2 H'' - \frac{w'}{w}H^2 H' \Big) (1+z)^2\nonumber \\
&&- \Big[2 (5 + 3w) H^2 H' - 3 \frac{w'}{w} H^3\Big] (1+z)\nonumber \\
&&+ 9  (1+w) H^3,
\end{eqnarray}

which if divided by $H_0^3$, can be expressed as 

\begin{eqnarray}
-wq  &=& 2 \Big(E E'^2 + E^2 E'' - \frac{w'}{w} E^2 E' \Big) (1+z)^2\nonumber \\ 
&&- \Big[ 2(5 + 3 w)E^2 E' - 3 \frac{w'}{w} E^3\Big](1+z)\nonumber \\
&&+ 9(1 + w)E^3,
\end{eqnarray}  

where $q = Q/H_0^3$ and $E = H/H_0$. 
%%%%%%%%%%%%%%%%%%%%%%%%%%%%%%%%%%%%%%%%%%%%%%%%%%

% Don't change these lines
\bsp	% typesetting comment
\label{lastpage}
\end{document}